\documentclass[a4paper,11pt]{article}
\pdfoutput=1 

\usepackage{jcappub} 

\usepackage[T1]{fontenc} 

\title{\boldmath Impossibility of rips and cosmological singularities in a universe merging with baby universes}

\author[a,1]{Oem Trivedi,\note{Corresponding author.}}
\author[b,c,d,2]{Maxim Khlopov}

\affiliation[a]{International Centre for Space and Cosmology, Ahmedabad University, \\ Ahmedabad 380009, India }
\affiliation[b]{Institute of Physics, Southern Federal University,\\ Stachki 194 Rostov on Don 344090, Russia }
\affiliation[c]{Virtual Institute of Astroparticle Physics, Paris 75018, France}
\affiliation[d]{National Research Nuclear University ”MEPHI”,\\ Moscow 115409, Russia}
\emailAdd{oem.t@ahduni.edu.in}
\emailAdd{khlopov@apc.in2p3.fr }

\abstract{Understanding the late-time acceleration of the universe and its subtleties is one of the biggest mysteries in cosmology. A lot of different approaches have been put forward to deal with this, ranging from the conventional cosmological constant to various models of dark energy and beyond. Recently one very interesting approach to explaining the late time acceleration has been put forward, where the expansion of the universe is driven by mergers with other "baby" universes and has been shown to be quite viable as well from the point of view of recent observational data. So in this work we examine the possibility of various rip scenarios and other future cosmological singularities in such "multiversal" scenario, probing such singularities for the first time in a multi universe scenario. We examine two models of such a baby universe merging cosmology, and show that remarkably no rip scenario or future cosmological singularity is possible in such models.}

\begin{document}
\maketitle
\flushbottom

\section{Introduction }
The surprising revelation of the late-time acceleration of the Universe posed a significant challenge for the field of cosmology \cite{SupernovaSearchTeam:1998fmf}. Since this discovery, extensive investigations have been undertaken to unravel the mystery behind this unexpected expansion phenomenon. Scientists have approached the cosmological expansion problem from diverse perspectives, employing traditional methods such as the Cosmological constant \cite{Weinberg:1988cp,Lombriser:2019jia,Padmanabhan:2002ji} and venturing into more unconventional theories like Modified Gravity \cite{Capozziello:2011et,Nojiri:2010wj,Nojiri:2017ncd}. Additionally, scenarios involving scalar fields driving late-time cosmic acceleration have been explored \cite{Zlatev:1998tr,Tsujikawa:2013fta,Faraoni:2000wk,Gasperini:2001pc,Capozziello:2003tk,Capozziello:2002rd,Odintsov:2023weg}. Quantum gravity approaches have also played a role in addressing the cosmic acceleration puzzle, spanning from Braneworld cosmology in string theory to theories like loop quantum cosmology and asymptotically safe cosmology \cite{Sahni:2002dx,Sami:2004xk,Tretyakov:2005en,Chen:2008ca,Fu:2008gh,Bonanno:2001hi,Bonanno:2001xi,Bentivegna:2003rr,Reuter:2005kb,Bonanno:2007wg,Weinberg:2009wa}. However, these endeavors have uncovered certain inconsistencies, underscoring the limitations of our current understanding of the universe. One of the most prominent challenges is the Hubble tension, indicating discrepancies in the values of the Hubble constant derived from detailed Cosmic Microwave Background (CMB) maps, combined with Baryon Acoustic Oscillations data and Supernovae Type Ia (SNeIa) data \cite{Planck:2018vyg,riess2019large,riess2021comprehensive,DiValentino:2021izs}. Therefore, the present epoch of the universe has presented us with a broad spectrum of questions and appears poised to become a domain where advanced gravitational physics will pave the way for a deeper comprehension of cosmology.
\\
\\
Various approaches to Quantum gravity have come forward in recent times and more often than not, they have also been shown to be of some insight into tackling the issues of late time cosmology and one such exciting approach to quantum gravity is that of Causal dynamical Triangulations or CDTs, which represent curved spacetimes with Lorentzian signature and serve as regularized lattice configurations in contemporary lattice gravity models \cite{Hamber:2009mt,Ambjorn:1998xu,Ambjorn:2000dv,Ambjorn:2001cv,Ambjorn:2004qm,Ambjorn:2004pw,Ambjorn:2005db,Ambjorn:2005qt}
. The fundamental idea behind CDT is to formulate a quantum gravity theory by taking a suitable scaling limit of a lattice theory. The dynamics of this lattice theory involve a nonperturbative path integral over geometric lattices, providing a direct approximation of curved spacetimes from classical general relativity. Similar to lattice QCD, the path integral relies on bare coupling constants and a UV lattice cutoff (denoted as 'a') representing the shortest length unit on the lattice.

As the lattice spacing 'a' is removed ($a \to 0$), one seeks continuum limits near a critical point where a certain correlation length in lattice units diverges. Achieving these continuum limits typically requires fine-tuning and renormalization of dimensionless bare couplings in the lattice model. 

For pure gravity within the CDT framework, the path integral $Z^{\rm CDT}$ takes the form of a continuum limit of a regularized lattice expression. Schematically, it is given by:
\[ Z^{\rm CDT}(G_{\rm N},\Lambda) = \lim_{{a\rightarrow 0}} \sum_{T\in{\cal T}} \frac{1}{C(T)} \, {\rm e}^{i S^{\rm CDT}[T]}, \]
where the sum is over inequivalent Lorentzian triangulations $T$, following specific causal gluing rules. Each term in the sum contributes with an amplitude determined by a lattice discretization $S^{\rm CDT}$ of the continuum Einstein-Hilbert action, given by:
\[ S^{\rm EH} = \frac{1}{G_{\rm N}} \int d^4x \, \sqrt{-\det g} \, (R[g,\partial g,\partial^2g]-2\Lambda), \]
where $G_{\rm N}$ and $\Lambda$ represent Newton's constant and the cosmological constant, respectively, and $R$ denotes the Ricci scalar of the metric tensor $g$. The quantity $C(T)$ in the expression for $Z^{\rm CDT}$ is the number of elements in the automorphism group of the triangulation $T$. This number is equal to 1 when $T$ lacks any symmetries, which is the typical scenario. The chosen version for the lattice action, denoted as $S^{\rm CDT}$, often follows the Regge version of the gravitational action, but this choice is not fundamental and is akin to a discrete approximation of general relativity. 
\\
\\
Another interesting thing is that Monte Carlo simulations have also been employed to explore the properties of the CDT model. Despite being defined for Lorentzian signature, numerical methods are applicable due to the presence of a well-defined Wick rotation—a rarity in quantum gravity beyond perturbation theory. One of the outcomes of these simulations reveals that when the spatial topology is $T^3$, the effective action as a function of the three-volume $V(t)$ at proper time $t$ is given by \cite{Ambjorn:2021chc,Ambjorn:2022wro}:
\begin{equation}\label{j1}
 S=  \frac{1}{\Gamma}\int dt  \, \Big( \frac{\dot{V}^2}{2 V} + \Lambda  V\Big).
\end{equation}
This finding is noteworthy as it essentially mirrors the Hartle-Hawking minisuperspace action, including the rotation of the conformal factor . The rotation of the conformal factor was proposed as a solution to the problem of the unboundedness from below of the Euclidean Einstein-Hilbert action. In CDT, the effective action arises through the integration (via Monte Carlo simulations) of all other degrees of freedom than $V(t)$, while Hartle and Hawking manually constrained the geometry to depend solely on $V(t)$.

Recalling the standard minisuperspace approximation, where the metric is given by:
\begin{equation}\label{j3}
 ds^2 = -N^2(t)dt^2 + a^2(t) d\Omega_3,\qquad d \Omega_3 = \sum_{i=1}^3 dx_i^2.
\end{equation}
We introduce:
\begin{equation}\label{j3a} 
 v(t) = \frac{1}{\kappa} a^3(t),\quad \kappa = 8\pi G,
\end{equation}
where $G$ is the gravitational constant. The minisuperspace Einstein-Hilbert action becomes:
\begin{equation}\label{j2}
 S =  \int dt  \,\Big(-\frac{\dot{v}^2}{3Nv} - \lambda N v\Big),
\end{equation}
with $\lambda$ being the cosmological constant. The Hamiltonian corresponding to \eqref{j2} is 
\begin{equation}\label{j4}
\mathcal{H}(v,p)  = N v \left(- \frac{3}{4} p^2+ \lambda \right),    
\end{equation}
where $p$ denotes the momentum conjugate to $v$. 
The analytic continuation of \eqref{j2} by Hartle and Hawking, including the rotation of the conformal factor, results in the following action:
\begin{equation}\label{jx2}
 S_{hh} =  \int dt  \,\Big(\frac{\dot{v}^2}{3Nv} + \lambda N v\Big).
\end{equation}
Hartle and Hawking primarily aimed to use \eqref{jx2} in the path integral. Regardless of the obtained result, one would eventually need to rotate back to Lorentzian signature for cosmological applications. If the quantum theory has a classical limit, this limit should be given by \eqref{j2} and \eqref{j4}. Thus, we do not anticipate the late-time aspects of cosmology to be directly impacted by the quantum aspects of gravity. Similarly, since CDT is designed as a quantum gravity theory, the result \eqref{j1} suggests that CDT may not offer new insights into late-time cosmology. "Traditional" quantum gravity might influence the early-time universe, addressing Big Bang singularities, etc., but not the late-time universe. However, if we allow for more "untraditional" quantum phenomena, such as the absorption and emission of so-called baby-universes \footnote{for more works on baby universes \cite{hawking1990baby,hawking1988baby,strominger1991baby,dijkgraaf2006baby,visser1990wormholes,hebecker2018euclidean,balasubramanian2020spin,marolf2021page,marolf2020transcending,giddings1989baby,hamada2022baby,maso2023birth} }, this situation can change as shown in \cite{Ambjorn:2023hnt}. So in this work we shall explore this scenario and shall see whether or not one can obtain rips and other cosmological singularities as we know them, in a universe where such a continual merger of other universes is happening as well. Essentially we are trying to tackle the issues of rips and singularities in the context of a Multiversal theory, which does seem to us as the first such endeavour in this direction. In the next section we shall describe this baby universe merging scenario in more detail yet briefly, while in section III we shall discuss the status of rips and other singularities in such a cosmology. We shall conclude our work in section IV. 
\\
\\
\section{Cosmic expansion with universe merger}
Here we shall briefly overview the scenario where late time expansion of the universe is through merging with other baby universes. Note that even though the word baby universe has been used here, it does not necessarily mean that the other universes are very small compared to ours but we use such nomenclature to be in line with the work in \cite{Ambjorn:2023hnt}, which we are following here. Given our consideration for other universes merging with our Universe, we are essentially discussing a multi-universe theory. Similar to a many-particle theory, it is natural to introduce creation and annihilation operators $\Psi^\dagger(v)$ and $\Psi(v)$ for single universes with spatial volume $v$. In a complete theory of four-dimensional quantum gravity, the spatial volume alone would not fully characterize a state at a given time $t$. Here, we simplify drastically by working in a minisuperspace approximation, where the spatial universe is entirely characterized by the spatial volume $v$. Thus, we denote the quantum state of a spatial universe with volume $v$ as $| v \rangle$. Considering the multi-universe Fock space constructed from these single universe states, we denote the Fock vacuum state as $| 0 \rangle$. The following commutation relation and operator actions apply:
\begin{align}\label{j11a}
 [\Psi(v),\Psi^\dagger(v')] &= \delta(v-v'), \\
 \Psi^\dagger(v) | 0\rangle &= | v\rangle, \\
 \Psi(v) |0\rangle &= 0.
\end{align}

In this manner, the (minisuperspace) quantum Hamiltonian, incorporating the creation and destruction of universes, can be expressed as:
\begin{align}\label{j12}
\hat{H} &= \hat{H}^{(0)} - g \int dv_1 \int dv_2 \;\Psi^\dagger(v_1)\Psi^\dagger(v_2)\;(v_1+v_2)\Psi(v_1+v_2) - \\
& g\int dv_1 \int dv_2 \;\Psi^\dagger(v_1+v_2)\;v_2\Psi(v_2)\;v_1\Psi(v_1) - \int \frac{dv}{v}\,  \rho(v) \Psi^\dagger (v), \nonumber
\end{align}
\begin{equation}\label{j13}
\hat{H}^{(0)} = \int_0^\infty \frac{dv}{v} \; \Psi^\dagger (v) \hat{\mathcal{H}}^{(0)} \, v  \Psi(v),~~~~~\hat{\mathcal{H}}^{(0)}=  v\left(-  \frac{3}{4} \frac{d^2}{dv^2}\,+\lambda  \right).
\end{equation}

Here, $\hat{H}^{(0)}$ describes the quantum Hamiltonian corresponding to the action \eqref{j1} (with $\beta = 2/3$), portraying the propagation of a single universe and note that we are taking $g> 0$ in our work \footnote{Although g=0 also form a class of solutions, which correspond to the simple dS space-times \cite{Ambjorn:2023hnt}.}. The two cubic terms describe the splitting of a universe into two and the merging of two universes into one, respectively. The last term suggests that a universe can be created from the Fock vacuum $| 0 \rangle$ provided the spatial volume is zero. Without this term, $\hat{H} | 0 \rangle = 0$, and the Fock vacuum would be stable. In our minisuperspace approximation, we do not attempt to describe how such merging or splitting realistically takes place; our primary interest lies in how the volume of space can be influenced by such processes, where the minisuperspace model provides valuable insights.

Even the minisuperspace Hamiltonian $\hat{H}$ is too complex to be solved in general. A universe can successively split and be joined by many others, and a part that splits off can later rejoin, thereby changing the topology of spacetime. The Hamiltonian is essentially dimension-independent (dimension dependence is absorbed in the coupling constants $\kappa$, $\lambda$, and $g$). Here we are dealing with string theory of 2 dimensional CDT and there exists a truncation that can be solved analytically of this called generalized CDT (GDCT). This theory has our main interest from a cosmological point of view as it follows the evolution of our Universe over time, considering the merging with other universes created at various times. After some analysis, we obtain:
\begin{equation}\label{j38}
\mathcal{H} =  v\Big( - \,\frac{3}{4}( p^2 + {\lambda} -2 g F( p) ) \Big) = \frac{3}{4} {v} \Big( (p +\alpha)  \sqrt{ ( p-\alpha)^2 +\frac{2 g}{\alpha} }\;\;\Big),
\end{equation}
where $p$ is the classical momentum conjugate to $v$. It was then shown that the Hamiltonian in \eqref{j38} can give the expansion of the universe without any cosmological constant, the expansion instead being driven by mergers with other universes. One can then include matter in the Hamiltonian, after which it can take the form
\begin{equation} \label{j30}
 \mathcal{H} [v,p] =  {v} \,( -f( p)+ \kappa \rho_{\rm m}(v) \,), 
 \quad v \rho_{\rm m}(v) = v_{pt} \rho_{\rm m} (v_{pt}), 
\end{equation}
where $v_{pt}$ and $\rho_{ m}(v_{pt})$ denote the values at the present time $t_{pt}$ The equations of motion for arbitrary $f(p)$ in \eqref{j30} can be written as
\begin{equation} \label{teq}
\dot{v} = \frac{\partial \mathcal{H}}{\partial p} = -v f'( p),\quad {\rm i.e.} \quad  3 \,\frac{\dot{a}}{a} =
\frac{\dot{v}}{v} = -f'( p)
\end{equation}
\begin{equation} \label{j42}
~~\dot{ p} = - \frac{\partial \mathcal{H}}{\partial v } =  f( p), \quad {\rm i.e.} \quad 
t- t_{\rm 0} = \int_{ p_{\rm 0}}^{ p} \frac{d \,  p}{f( p)}
\end{equation}
$p_{0}$ is p at some initial time $t_{0}$, which is not necessarily the start of the universe here and the primes denote differentiation with respect to p. 
From \eqref{teq} we see that the Hubble parameter can be written as \begin{equation} \label{heq}
    H = - \frac{f^{\prime} (p)}{3}
\end{equation}
We are now at a point from which we can slowly transition to analyzie rips and singularities, which we shall now do in the next section. 
\section{Impossibility of rips and singularities in this "Multiverse"}
In recent times, a substantial body of literature has emerged, focusing on the exploration of various types of singularities that may arise in the present and distant future of the Universe. The detection of late-time acceleration has significantly propelled such investigations \cite{Nojiri:2004ip,Nojiri:2005sr,Nojiri:2005sx,Bamba:2008ut,trivedi2022finite,trivedi2022type,odintsov2015singular,odintsov2016singular,oikonomou2015singular,nojiri2015singular,odintsov2022did,Trivedi:2023aes,Trivedi:2023rln,Trivedi:2023wgg,Trivedi:2023zlf,deHaro:2023lbq,Colgain:2021beg,Odintsov:2023qfj,Brevik:2021wzs,Nojiri:2022xdo,Odintsov:2021yva,Odintsov:2022unp,Battista:2020lqv}.
 The term "singularity" encompasses diverse definitions, and until the late 20th century, the primary instances of singularity formation in cosmology were the initial Big Bang singularity and, in spatially closed cosmological models, the eventual Big Crunch singularity. Hawking and Penrose provided the definition of a singular point in cosmology, and many of the theorems they established rely on the null energy condition. Moreover, at a singular point in spacetime, geodesic incompleteness and the divergence of curvature scalars occur. Even though modified gravity might alter the null energy condition compared to the Einstein-Hilbert case, it is widely acknowledged that geodesic incompleteness and the divergence of curvature invariants strongly suggest the existence of a crushing singularity.
\\
\\
The effects of singularities in cosmology vary, and a comprehensive classification was conducted in \cite{Bamba:2008ut,capozziello2009classifying}. While one can interpret singularities as points where a cosmological theory encounters challenges, they can also be seen as gateways to new physics, adding a different kind of intriguing interest. In particular, finite-time singularities (those occurring within a finite time) could be perceived as either imperfections in classical theory or as portals to a quantum description of general relativity. This distinction arises because these singularities cannot be treated similarly to the spacelike singularities of black holes. Therefore, questions about the accuracy of predictions made by classical gravitational theories arise. Consequently, the study of singularities in cosmological contexts and exploring methods to potentially eliminate them contribute to a deeper understanding of the connection between quantum descriptions of cosmology and their classical counterparts. The rip scenarios and singularities we are talking of are \begin{itemize}
    \item Big rip (Type I singularity) : A well known scenario, where for $t \to t_{f}$, where $t_{f}$ is finite, we have both the effective energy density and pressure density of the universe diverging, $p_{eff} \to \infty, \rho_{eff} \to \infty $, while we also have a diverging Hubble parameter $H \to \infty$ \cite{Caldwell:2003vq}. This results in a scenario of universal death, where everything within the universe undergoes progressive disintegration. \cite{Caldwell:2003vq}.
    \item Sudden/Pressure singularity (Type II singularity): In this case, $p_{\text{eff}}$ diverges, as well as the derivatives of the scale factor beyond the second derivative \cite{Barrow:2004xh,Andersson:2000cv} (Big Brake singularities are a special case of this \cite{Gorini:2003wa} ).
    \item Big Freeze singularity (Type III singularity): In this case, the derivative of the scale factor from the first derivative onwards diverges. These were detected in generalized Chaplygin gas models \cite{bouhmadi2008worse}.
    \item Generalized sudden singularities (Type IV singularity): These are finite time singularities with finite density and pressure instead of diverging pressure. In this case, the derivative of the scale factor diverges from a derivative higher than the second \cite{Bamba:2008ut,Nojiri:2004pf} .
    \item Little rip : It is similar to the big rip, but here where for $t \to \infty$, we have the effective energy density, pressure density and Hubble parameter diverging, $p_{eff} \to \infty, \rho_{eff} \to \infty $, $H \to \infty$. It is effectively a big rip which happens at infinite time \cite{Frampton:2011sp}. 
    \item Pseudo rip : The Pseudo rip also takes place at finite time, but in this case we have a finite Hubble parameter, so $H \to H_{f} $ for a finite $H_{f}$ as $t\to \infty$ \cite{Frampton:2011aa}
    \item Little sibling of the Big rip : In this case at infinite times, we have a diverging Hubble parameter again just like the Little rip but the derivatives of the Hubble parameter are finite \cite{Bouhmadi-Lopez:2014cca}. 
\end{itemize}

We are interested to see which of these scenario can be allowed for in the models prepared in \cite{Ambjorn:2023hnt}. The interest here stems from the very exciting notion of this universe, where smaller baby universes are merging into it and so seeing what possibilities it could have in the future realistically seems to be something very worth knowing. 
\\
\\
With the singularities discussed in detail, we need to know inculcate the pressure and energy terms in our formulation and this can be done in a similar way to usual cosmology as achieved in \cite{Ambjorn:2023hnt}. One can write various important cosmological parameters in terms of p using the formulation as discussed in section II, for example Redshift, angular diameter etc. but what we shall be interested in here are the so called "formal pressure" and "formal energy", which are the equivalents of the pressure and energy densities in this cosmology.  Any solution to \eqref{teq}-\eqref{j42} is inherently subject to the condition $\mathcal{H} = {\rm const}$ by construction. Our focus lies on the "on-shell" solutions where $\mathcal{H} = 0$, implying that
\begin{equation}\label{j40}
    f(p) = \kappa \rho_{\rm m}(v) = \kappa \rho_{\rm m}(v_{tp}) \frac{v_{tp}}{v} = f(t_{tp}) \frac{t_{tp}}{v},
\end{equation}
where $p_0$ represents the value of $p$ at the present time $t_{tp}$, with \eqref{j40} being recognized as the "Generalized Friedmann equation" \cite{Ambjorn:2023hnt}.  The "formal density" $\rho_f(t)$ or $\rho_f(p)$ associated with the function $f(p)$ is obtained by expressing the generalized Friedmann equation as
\begin{equation}\label{def3}
    \left(\frac{\dot{a}(t)}{a(t)}\right)^2 = \frac{\kappa \rho_{\rm m}(v)}{3} + \frac{\kappa \rho_f(v)}{3},
\end{equation}
from which, utilizing the equations of motion, we derive
\begin{equation}\label{28}
    \kappa \rho_f(p) = \frac{1}{3} \left(f'(p)\right)^2 - f(p).
\end{equation}
\begin{equation}\label{29}
    \kappa \frac{d \rho_f}{dt} = f(p) f'(p) \left(\frac{2}{3} f''(p) - 1\right).
\end{equation}
We introduce the "formal pressure" $P_f$ associated with $\rho_f$ through the energy conservation equation
\begin{equation}\label{def6}
    \frac{d}{dt} (v \rho_f) + P_f \frac{d}{dt} v = 0.
\end{equation}
This leads to
\begin{equation}\label{31}
    P_f = f(p) \left(\frac{2}{3} f''(p) - 1\right) - \rho_f(v),
\end{equation}
and the "formal equation of state parameter" $w_f$ is defined (for $\rho_f \neq 0$) as
\begin{equation}\label{def8}
    w_f = \frac{P_f}{\rho_f} = \frac{f(p) \left(\frac{2}{3} f''(p) - 1\right)}{\frac{1}{3} \left(f'(p)\right)^2 - f(p)} - 1.
\end{equation}
One also hence notes that \begin{equation} \label{correctp}
     P_{f} = \frac{1}{\kappa} \Bigg[ f(p) \left( \frac{2 f''(p}{3} -1 \right) + f(p) - \frac{f'(p)^2}{3} \Bigg]  
\end{equation}
The definitions of $\rho_f$ and $P_f$ ensure that our equations of motion can be expressed in the standard GR form, using $a$, $\rho_{\rm m}$, $\rho_f$, and $P$. Till now we have maintained the notation used in \cite{Ambjorn:2023hnt} for formal pressure and formal density, but from here onwards formal pressure would refer to $\Tilde{p}$ and formal density to $\Tilde{\rho}$ in this work. Furthermore, differentiating \eqref{heq} with respect to time, and using \eqref{teq} in doing so, we then arrive at \begin{equation} \label{hint}
    \int_{H_{0}}^{H_{f}} \frac{dH}{H} = \int_{p_{0}}^{p_{f}} \frac{f^{\prime \prime} (p)}{ f^{\prime} (p)} dp
\end{equation} 
Dividing \eqref{29} by \eqref{28} and again using \eqref{teq}, we arrive at the following for the formal density \begin{equation} \label{eneeq}
    \int_{\Tilde{\rho}_{0}}^{\Tilde{\rho}_{f}} \frac{d \Tilde{\rho}}{\Tilde{\rho}} = \int_{p_{0}}^{p_{f}} \frac{ f^{\prime} (p) \left( 2 f^{\prime \prime} (p) - 3 \right) }{{f'(p)}^2 - 3 f(p) } dp
\end{equation}
Similarly, taking the derivative of the formal pressure with respect to p using \eqref{correctp} we can write \begin{equation} \label{preseq}
\int_{\Tilde{p}_{o}}^{\Tilde{p}_{f}} \frac{d \Tilde{p}}{\Tilde{p}} = \frac{2 f(p) f'''(p)+f'(p) \left(6-4 f''(p)\right)}{2 f(p) \left(f''(p)+3\right)-3 f'(p)^2}
\end{equation} 
We can also write for the derivatives of H as \begin{equation} \label{dotheq}
    \int_{\Dot{H}_{0}}^{\Dot{H}_{f}} \frac{d\Dot{H}}{\dot{H}} = \int_{p_{0}}^{p_{f}} \frac{f^{\prime} (p) f^{\prime \prime} (p) + f(p) f^{\prime \prime \prime} (p) }{f^{\prime \prime} (p) f(p)} dp
\end{equation}
\begin{equation} \label{ddotheq}
    \int_{\Ddot{H}_{o}}^{\Ddot{H}_{f}} \frac{d\Ddot{H}}{\ddot{H}} = \frac{{f^{\prime} (p)}^2 f^{\prime \prime } (p) + f(p) {f^{\prime \prime }(p)}^2 + f(p)^2 f^{\prime \prime \prime \prime } (p) + 3 f^{\prime \prime }(p) f(p) f^{\prime \prime \prime}(p)}{f(p) \left( f^{\prime} (p) f^{\prime \prime} (p) + f(p) f^{\prime \prime \prime} (p) \right)} dp    
\end{equation}
One notes that our results will be invariant under the transformations $\kappa \to \frac{\kappa}{\lambda} $, $\Tilde{\rho} \to \lambda \Tilde{\rho}$, $\Tilde{p} \to \lambda \Tilde{p}$, where $\lambda$ is some constant, as suggested by the characteristics of the equations \eqref{28}-\eqref{correctp}. Now we can identify the realism of the various rip scenarios and future singularities we have outlined above as follows \begin{itemize}
    \item Big rip (type I) : We firstly check if there is a $p_{f}$ for which the RHS in \eqref{hint} diverges, and if we do find such, we then apply that p in \eqref{teq} to check if the LHS in it is non divergent. If both conditions are satisfied, alongside having diverging RHS in \eqref{eneeq} and \eqref{preseq}, then we can say that for that particular $f(p)$, a big rip is a realistic scenario.
    \item Pressure / Big Brake/Sudden singularities (Type II) : In this case we would just have to check if there is a $p_{f}$ for which \eqref{preseq} diverges, while others remain finite.
    \item Big Freeze (Type III) : In this case we would have to check if there is a $p_{f}$ for which \eqref{preseq} and \eqref{eneeq} diverge, but \eqref{hint} and \eqref{teq} do not.
    \item Generalized sudden singularities (Type IV) : In this case, we would have to check if the RHS in \eqref{ddotheq} diverges for certain $p_{f}$ while other integrals do not.
    \item Little rip and Pseudo rip: We firstly check if there is a $p_{f}$ for which the LHS in \eqref{teq} diverges, and if we do find such, we then apply that p in \eqref{hint} to check if the RHS in it diverges or not. We would also check whether the RHS in \eqref{eneeq} and \eqref{preseq} diverges as well. If the RHS in \eqref{hint} diverges, we have a Little rip in this scenario while if it is non divergent, we then have a Pseudo rip for that particular $f(p)$
    \item Little sibling of the big rip : In this case we would have to check whether the RHS in \eqref{dotheq} diverges or not for the p for which $t \to \infty$, besides the checks as above. 
\end{itemize} 
We now need to consider particular models of baby universe merger driven cosmic expansion to fully understand what we want to and for that we need to finally specify $f(p)$ and we will focus on the baby universe merging models provided in \cite{Ambjorn:2023hnt}, those being \begin{equation} \label{fgcdt}
    f_{gcdt} (p) = -\frac{3}{4} (p+\alpha) \sqrt{(p-\alpha)^2 + 2 \alpha^2} 
\end{equation}
and \begin{equation} \label{fmod}
    f_{mod} (p) = \frac{3}{4} \left(  p^2 + \frac{2g}{p} \right) 
\end{equation}
where $\alpha = g^{1/3} $ and $g \geq 0$
We will firstly talk of the model in \eqref{fmod}. For this model it was already shown in \cite{Ambjorn:2023hnt} that $t \to \infty , p-> - 2^{1/3} \alpha$, so that is the highest value any p can take, which gives us a constraint $$p_{f} \leq - 2^{1/3} \alpha $$ This is a negative value as $g \geq 0$, in fact it was shown in \cite{Ambjorn:2023hnt} how $g=0$ just corresponds to the simple dS solution, which they also discussed a toy model in their paper. But for actually relevant purposes from the point of a view of a cosmos with merging universes, it is imperative to take $g>0$, which was again showed to still give an expanding universe even without a cosmological constant. 
\\
\\
Now, we calculate the relevant integrals \eqref{hint},\eqref{eneeq}, \eqref{preseq}, \eqref{dotheq}, \eqref{ddotheq} in this case, starting with \eqref{hint} \begin{equation} \label{hf3}
    \int_{H_{0}}^{H_{f}} \frac{dH}{H} = \ln \left(\frac{(p_{f}^3-g) p_{0}^2}{p_{f}^2(p_{0}^3-g)} \right)
\end{equation}
\begin{equation} \label{prf3}
  \int_{\Tilde{p}_{0}}^{\Tilde{p}_{f}} \frac{d \Tilde{p}}{\Tilde{p}} =  \ln \left( \frac{(g + 14 p_{f}^3) p_{0}^4}{(g + 14 p_{0}^3) p_{f}^4} \right)
\end{equation}
\begin{equation} \label{rf3}
    \int_{\Tilde{\rho}_{0}}^{\Tilde{\rho}_{f}} \frac{d \Tilde{\rho}}{\Tilde{\rho}} = \ln \left(\frac{(4 p_{f}^3-g) p_{0}^4}{p_{f}^4 (4 p_{0}^3-g)}\right) 
\end{equation}
\begin{equation} \label{dhf3}
    \int_{\dot{H}_{o}}^{\dot{H}_{f}} \frac{d \dot{H}}{\dot{H}} = \ln \left(\frac{(p_{f}^3 + 2g)^2 p_{0}^4}{p_{f}^4 (p_{0}^3 + 2g)^2}\right) 
\end{equation}
\begin{multline} \label{ddhf3}
    \int_{\ddot{H}_{o}}^{\ddot{H}_{f}} \frac{d \ddot{H}}{\ddot{H}} = \frac{3 \sqrt[3]{2}}{\sqrt[3]{g}} \Bigg[\ln \left(\frac{4 g^{2/3}+2 \sqrt[3]{2} \sqrt[3]{g} p_{f}+2^{2/3} p_{f}^2}{4 g^{2/3}+2 \sqrt[3]{2} \sqrt[3]{g} p_{0}+2^{2/3} p_{0}^2}\right)+\ln \left(\left(\frac{2 \sqrt[3]{g}-\sqrt[3]{2} p_{0}}{2 \sqrt[3]{g}-\sqrt[3]{2} p_{f}}\right)^2\right) \\ +2 \sqrt{3} \left(\tan ^{-1}\left(\frac{\frac{\sqrt[3]{2} p_{0}}{\sqrt[3]{g}}+1}{\sqrt{3}}\right)-\tan ^{-1}\left(\frac{\frac{\sqrt[3]{2} p_{f}}{\sqrt[3]{g}}+1}{\sqrt{3}}\right)\right)\Bigg]  +\ln \left(\left(\frac{p_{f}^3-4 g}{p_{0}^3-4 g}\right)^{14}\right)+\ln \left(\left(\frac{2 g+p_{f}^3}{2 g+p_{0}^3}\right)^4\right) \\ +\frac{36 (p_{f}-p_{0})}{p_{0} p_{f}}+\ln \left(\left(\frac{p_{0}}{p_{f}}\right)^{30}\right)
\end{multline}
 In \cite{Ambjorn:2023hnt} it was considered that $p_{0} \to - \infty $ but we here are primarily only concerned with the properties of $p_{f}$ so we do not need to specify any $p_{0}$ as well. 
\\
\\
We firstly note the impossibility of the big rip in this scenario, as the integral \eqref{hf3} would not diverge for any allowed $p_{f}$ as it can only diverge for a value  $p_{f} \to 0$, which is unrealistic considering that p cannot have non-negative values. It is important to realize that $p_{f}$ cannot attain positive values or zero, as we have taken $g > 0 $ and time reaches infinity as $p \to - 2^{1/3} \alpha$. So going to values like 0 or positive for p would mean going "beyond infinity" in some way, which does not make sense here as well. So that would mean H would not diverge in \eqref{hf3}. Similarly, we see that $\Tilde{p}$, $\Tilde{\rho}$ and $\dot{H}$  would also not diverge for any value of p, as all the expressions in \eqref{prf3},\eqref{rf3} and \eqref{dhf3} would only diverge for $p \to 0 $, which is not a possibility as we have seen here. This would mean that we do not have the possibilities of the Little rip and Pseudo rip either, the energy density and pressure density does not diverge in any case. The Little sibling of the big rip is not a possibility as well as H does not diverge at infinite times. The Sudden singularity ( type II ) and Big Freeze singularity (type III) are not possible as well hence. For type IV singularity, we see that $\Ddot{H}$ can diverge as in \eqref{ddotheq} for $p \to 0 $ and also for $p \to 2^{2/3} g = 2^{2/3} \alpha $. But we again note that p is constrained as $p \leq - 2^{1/3} \alpha $, which means that the value of p for which the EOS parameter diverges is not attainable and hence a type IV singularity is also not possible.  We now come to the very intriguing conclusion that in such a cosmic scenario where baby universes keep continually merging with our universe, we do not observe any future cosmological singularities or at least, not the ones we are aware of. This is incredibly exciting, as this is the first time the question of cosmological singularities has been addressed from the point of view of a multiverse and seeing that all the prominent cosmological singularities do not exist in a multiversal setting is certainly bewildering. Furthermore seeing in particular how all the main rip scenarios are not possible in such a universe, it also makes one wonder how such a baby merging universe would eventually end.  
\\
\\
Now for the model \eqref{fgcdt}, we note that in this case, as shown in \cite{Ambjorn:2023hnt} it is not entirely clear for what value of $p$ $t \to \infty$, so in this case the Little rip, Pseudo rip and Little sibling of the big rip lie beyond our scope to begin with.  The integral \eqref{hint} reads as follows 
\begin{equation} \label{hgcdt}
    \int_{H_{o}}^{H_{f}} \frac{dH}{H} =  \ln \left(\frac{\alpha ^2+p_{f}^2-\alpha  p_{f}}{\alpha ^2+p_{0}^2-\alpha  p_{0}} \sqrt{\frac{3 \alpha ^2+p_{0}^2-2 \alpha  p_{0}}{3 \alpha ^2+p_{f}^2-2 \alpha  p_{f}}}  \right)
\end{equation}
 We note here that the integral above can only diverge for a complex value of p, to be precise for $p_{f} = \alpha \pm i \sqrt{2} \alpha$, which is of course unrealistic. Similarly, the integrals \eqref{preseq}, \eqref{eneeq} and \eqref{dotheq} read 
 \begin{equation} \label{dhgcdt}
    \int_{\dot{H}_{o}}^{\dot{H}_{f}} \frac{d \dot{H}}{\dot{H}} = \ln \left(\frac{(\alpha +p_{f}) \left(-2 \alpha ^3+p_{f}^3-3 \alpha  p_{f}^2+6 \alpha ^2 p_{f}\right) (3 \alpha ^2+p_{0}^2-2 \alpha  p_{0})}{(3 \alpha ^2+p_{f}^2-2 \alpha  p_{f} ) \left( (\alpha +p_{0}) \left(-2 \alpha ^3+p_{0}^3-3 \alpha  p_{0}^2+6 \alpha ^2 p_{0}\right) \right)  }\right) 
\end{equation}
\begin{equation} \label{prgcdt}
  \int_{\Tilde{p}_{0}}^{\Tilde{p}_{f}} \frac{d \Tilde{p}}{\Tilde{p}} =  \ln \left( \frac{r_{1} (p_{f}) r_{2} (p_{0}) }{r_{2} (p_{f}) r_{1} (p_{0})} \right)
\end{equation}
\begin{equation} \label{rgcdt}
    \int_{\Tilde{\rho}_{o}}^{\Tilde{\rho}_{f}} \frac{d \Tilde{\rho}}{\Tilde{\rho}} = \ln \left( \frac{q_{1}(p_{0}) q_{2}(p_{f})}{q_{1}(p_{f}) q_{2}(p_{0}) } \right)
\end{equation}
where 
$$ q_{1}(p) = \left(\alpha +\sqrt{3 \alpha ^2+p^2-2 \alpha  p}-p\right)^2 \left(3 \alpha ^2+(\alpha -p) \sqrt{3 \alpha ^2+p^2-2 \alpha  p}+p^2-2 \alpha  p\right)^2 $$ \begin{multline*}
    q_{2}(p) = 43 \alpha ^5-8 p^5+39 \alpha  p^4-98 \alpha ^2 p^3+139 \alpha ^3 p^2+ \\ \left(25 \alpha ^4+8 p^4-31 \alpha  p^3+59 \alpha ^2 p^2-57 \alpha ^3 p\right) \sqrt{3 \alpha ^2+p^2-2 \alpha  p}  -114 \alpha ^4 p     
\end{multline*}
\begin{multline*}
    r_{1}(p) = 107 \alpha ^5-28 p^5+132 \alpha  p^4-316 \alpha ^2 p^3+422 \alpha ^3 p^2+ \\ \left(62 \alpha ^4+28 p^4-104 \alpha  p^3+184 \alpha ^2 p^2-162 \alpha ^3 p\right) \sqrt{3 \alpha ^2+p^2-2 \alpha  p}-318 \alpha ^4 p 
\end{multline*}
\begin{multline*}
    r_{2}(p) = \left(\alpha +\sqrt{3 \alpha ^2+p^2-2 \alpha  p}-p\right)^2 \left(-3 \alpha ^2+(p-\alpha ) \sqrt{3 \alpha ^2+p^2-2 \alpha  p}-p^2+2 \alpha  p\right)^2
\end{multline*}
We again note that all the above integrals will only diverge for a complex values of p, which is the same as that for H which is $p_{f} = \alpha \pm i \sqrt{2} \alpha $. The expression for \eqref{ddotheq} is awfully long and so we dont write that here, but similarly it will also blow up only for this complex value. This is of course unrealistic again and so expecting these quantities to blow up is also outside our realm of possibilities. This tells us that there would be no finite time blow ups of the energy density, pressure density, Hubble parameter and its derivatives even up till the second order. This tells us that the big rip , Pressure/ Big Brake/ Sudden singularity, Big Freeze and generalized sudden singularities are all beyond the realm of possibilities in this model as well.
\\
\\
\section{Conclusions}
In this work, we have pondered about the realism of rips and other cosmological singularities in a universe which is continually merging with other universes and driving its expansion via these mergers. We firstly discussed how such universes can come into being with the exciting approach of Causal dynamical triangulations to quantum gravity, discussing the universal wavefunction approach in this scenario and other subtleties. We then formulated conditions on how we can attest to whether or not the rip scenarios (big rip, little rip, quasi rip, pseudo rip ) and other cosmological singularities could occur in such a universe, after which we analyzed this occurrence for two such viable multiversal models. The conclusion we see from analyzing the two models with baby universe mergers into our own universe in \eqref{fmod} and \eqref{fgcdt} is that such a multiversal scenario somehow gives us rise to a cosmos where we do not see the prominent future cosmological singularities taking place. It is especially interesting, because one might imagine that if one has new universes merging with another universe (in this case our own) then such blow ups may become more frequent but what we end up with is a way more refined cosmology. It is even more remarkable given the fact that this scenario develops out after considering just the Einstein-Hilbert Action in the framework of CDT, as the conventional EH action has been frequently shown to be riddled with these singularities and attempts have always been made to smoothen them out. Even though many such attempts, from quantum gravity perspectives like conformal anomalies or modified gravity effects, have been pursued one sees that not all the singularities can be delayed or avoided even in the most optimistic of cases. So this springs a huge surprise, as by considering a Multiversal scenario one ends up with a universe which is apparently free from the prominent cosmological singularities and rips. This may point towards the direction of the notion of the Multiverse having some sort of realism too, as the other papers on this topic \cite{Ambjorn:2021chc,Ambjorn:2022wro} have also shown that such a universe can accomodate large scale structure and also the Hubble tension ( particularly the model \eqref{fmod}). This also, to the best of our knowledge, is the first work ever which has considered such cosmological singularities in the context of a multiverse. It is also important to mention how this current work differs from previous treatments of the multiverse in this cosmological paradigm as in \cite{Ambjorn:2023hnt,Ambjorn:2017cxu} as while in these papers there have been certain mentions of singularities, a full and extensive analysis taking into account all kinds of future cosmological singularities has not been undertaken in any similar way to the one we have presented here. Our work has comprehensively taken the status quo of various cosmologically interesting parameters and have investigated their evolution towards future times, focusing on how any blowups in these parameters can be achieved and whether one can even have blowups in them, which we have ended up showing is not the case. These cosmological singularities carry special interest from the point of view of understanding the late time acceleration of the universe and so properly examining them in this paradigm, which also promises to be prospectively an interesting explanation for the late expansion, is something of paramount importance and so we have pursued this.
	
	\acknowledgments

The authors would like to thank Jan Ambjørn for helpful discussions on various aspects of CDT and baby universe mergers. The research by M.K. was carried out in Southern Federal University with financial support of the Ministry of Science and Higher Education of the Russian Federation (State contract GZ0110/23-10-IF). We would like to thank the editor/referee of the work for his/her comments on it, which have greatly improved the depth of the paper.


\bibliography{JSPJMJcitations.bib}

\bibliographystyle{unsrt}

\end{document}